\documentclass[11pt,a4paper,DIV=14]{scrartcl}
\usepackage{natbib}
\usepackage{amsmath,amsfonts,amssymb,graphicx,enumerate,color,verbatim,enumerate,xcolor}
\usepackage{moreverb}
\usepackage[colorlinks,bookmarksopen,bookmarksnumbered,citecolor=red,urlcolor=red]{hyperref}
\newcommand\BibTeX{{\rmfamily B\kern-.05em \textsc{i\kern-.025em b}\kern-.08em
T\kern-.1667em\lower.7ex\hbox{E}\kern-.125emX}}

\renewcommand{\vec}[1]{\boldsymbol{#1}}

\newcommand{\logit}{\text{logit}}
\newcommand{\elogit}{\text{elogit}}
\newcommand{\eexpit}{\text{eexpit}}
\newcommand{\thetavec}{\vec{\theta}}
\newcommand{\Thetavec}{\vec{\Theta}}

\newcommand{\nvec}{\vec{n}}
\newcommand{\rvec}{\vec{r}}
\newcommand{\tvec}{\vec{t}}
\newcommand{\yvec}{\vec{y}}
\newcommand{\xvec}{\vec{x}}
\newcommand{\pvec}{\vec{p}}
\renewcommand{\epsilon}{\varepsilon}
\newcommand{\LN}{\text{LN}}
\newcommand{\diag}{\text{diag}}

\usepackage{booktabs}
\usepackage{algorithm}
\usepackage{longtable}
\usepackage{setspace}
\usepackage{amssymb}
\usepackage{microtype}
\usepackage[british]{babel}

\renewcommand{\vec}{\boldsymbol}
\renewcommand{\hat}{\widehat}
\newcommand{\D}{\mathcal{D}}
\newcommand{\K}{\tilde{K}}

\graphicspath{{figs/}}

\newcommand\blfootnote[1]{%
  \begingroup
  \renewcommand\thefootnote{}\footnote{#1}%
  \addtocounter{footnote}{-1}%
  \endgroup
}
\begin{document}

%\runningheads{Inference for a partially observed process using data on proportions}{R. J. Boys, H. F. Ainsworth and C. S. Gillespie}

\title{Bayesian inference for a partially observed birth-death process
  using data on proportions}
\author{R. J. Boys$^{1}$\thanks{email: \texttt{richard.boys@ncl.ac.uk}}, H. F. Ainsworth$^2$ and C. S. Gillespie$^1$}

%\affiliation{University of Newcastle~upon~Tyne}

\date{$^1$School of Mathematics, Statistics \& Physics, Newcastle
  University, Newcastle~upon~Tyne,~NE1~7RU, UK.\\ $^2$Institute of
  Health and Society, Newcastle University,
  Newcastle~upon~Tyne,~NE1~4AX, UK.}

%\ack{This class file was developed by Sunrise Setting Ltd, Paignton, Devon, UK. Website:\\ \href{http://www.sunrise-setting.co.uk}{\texttt{www.sunrise-setting.co.uk}}}

\maketitle
\begin{abstract}
Stochastic kinetic models are often used to describe complex
biological processes. Typically these models are analytically
intractable and have unknown parameters which need to be estimated
from observed data. Ideally we would have measurements on all
interacting chemical species in the process, observed continuously in
time. However, in practice, measurements are taken only at a
relatively few time-points. In some situations, only very limited
observation of the process is available, such as when experimenters
can only observe noisy observations on the proportion of cells that
are alive. This makes the inference task even more problematic. We
consider a range of data-poor scenarios and investigate the
performance of various computationally intensive Bayesian algorithms
in determining the posterior distribution using data on proportions
from a simple birth-death process.
\end{abstract}
\blfootnote{Holly Ainsworth was supported by a PhD studentship from the UK
  Engineering and Physical Sciences Research Council.  The authors
  would like to thank an Editor and three anonymous referees for
  comments that improved the paper.}

%\footnotetext[2]{Please ensure that you use the most up to date class file, available from the ANZS Home Page at\\ \href{http://onlinelibrary.wiley.com/journal/10.1111/(ISSN)1467-842X}{\texttt{http://onlinelibrary.wiley.com/journal/10.1111/(ISSN)1467-842X}}}

\section{Introduction}
Biological modellers increasingly use stochastic kinetic models to
describe the complex and stochastic nature of their experiments.
Typically these models are analytically intractable and have unknown
parameters which need to be estimated from observed data. Ideally the
experiments would produce continuous-time measurements on all chemical
species within the model, and with such data, Bayesian inference
usually proceeds in a straightforward manner, often by taking
conjugate priors for the model parameters. However, in practice,
continuous-time measurements are not possible and measurements are
taken only at a relatively few time-points. This complicates the
inference as the observed data likelihood is typically intractable.
Solutions to this problem often use data augmentation and thereby
integrate over the unobserved continuous paths between observations;
see, for example, \cite{Gibson1998}, \cite{BWK08}, \cite{Gibson2001}
and \cite{GoliWilk05}. In some situations, experiments only yield a
very limited view of the underlying process. For example, all that may
be observed could be noisy observations on the proportion of cells
that are alive. This can make the inference task even more
problematic. In this paper we consider a range of data-poor scenarios
and build computationally intensive Bayesian algorithms to determine
the posterior distribution. We will study a system in which
measurements are available on the alive-status of individual cells and
describe the (independent) dynamics of each cell using a simple
birth-death process.  Cell death is assumed to occur when its
internal population (as described by the birth-death process) becomes
extinct. The birth-death process provides useful toy model which
captures a wide range of lifetime distributions for different choices
of its birth and death rates and initial population size.

In section~2, we describe the simple birth-death process and state
some of the key analytic expressions that will be needed to formulate
expressions for the likelihood function in the data scenarios we
consider. Section~3 goes on to describe these data scenarios and
outlines methods for obtaining the posterior distribution. In
section~4 we consider the case where we do not have an analytic
expression for the probability that the cell is dead at
time~$t$. Instead we have to base our inference scheme on simulated
proportions of cell death obtained from running a simulator of the
underlying (birth-death) process. In many cases, obtaining these
simulated proportions within an MCMC scheme will be far too computer
intensive and so we consider other methods which make use of Gaussian
process emulators.

\section{The birth-death process}
The simple birth-death process is a well studied stochastic
model. The model, which dates back to \cite{Yule1925} and
\cite{FellerW1939}, has been widely used in biological applications;
see, for example, its use as a model for the early stages of an
epidemic in \cite{Kendall1948}. Key useful attributes of the model are
its simplicity and tractability. The model for population size~$X$ in
a typical cell can be written in chemical notation as
$R_1:X\stackrel{\lambda}{\rightarrow}2X$ and
$R_2:X\stackrel{\mu}{\rightarrow}\emptyset$, where $\lambda$ and $\mu$
are the birth and death rates (per member of the population).
%, with hazard functions $h_1(x,\lambda)=\lambda x$ and $h_2(x,\mu)=\mu x$.

The model is sufficiently simple that it is possible to obtain an
analytic expression for its transition probabilities; see
\cite{Renshaw:1993}. In particular, for $\lambda\neq\mu$ and an
initial population of size $x_0$ within the cell, the probability of
cell death (i.e. population extinction within the cell) in $[0,t]$ is
\begin{equation}
P_0(t)=\left\{\frac{\mu-\mu e^{(\mu-\lambda)t}}{\lambda-\mu e^{(\mu - \lambda)t}}\right\}^{x_0}
\label{eq:bd_cdf}
\end{equation}
and the density of extinction by time~$t$ is
\begin{equation}
p_0(t) =
\frac{x_0\mu^{x_0} (\lambda-\mu)^2 e^{(\mu-\lambda)t}
\{1-e^{(\mu-\lambda)t}\}^{x_0}}{\{1-e^{(\mu-\lambda)t}\}\{\lambda -\mu e^{(\mu-\lambda)t}\}^{(x_0 + 1)}}.
\label{eq:p0}
\end{equation}
Due to the tractable nature of the process, many authors have used the
system as a test bed for different scenarios. For example,
\cite{Dehay2007} consider parameter inference when observing the
process at discrete equi-distant time points, while
\cite{Gillespie:2008} consider the case where only deaths are
observed; see also the references therein. Inference for more general
birth-death processes has also been considered by numerous authors;
see, for example, \cite{Crawford2012} and \cite{Crawford2014}. Also
\cite{bladt05} consider inference in discretely observed Markov jump
processes.

\section{Inference under various data poor scenarios}
We now consider three observational scenarios and outline how
realisations can be simulated from the parameter posterior
distribution using MCMC methods. The scenarios are
\begin{enumerate}[(a)]
\item the times of cell death are known (exactly), with data
  $\tvec^e=(t_1^e,t_2^e,\ldots,t_m^e)$;
\item the status of each cell is observed at time points $t_1,t_2,\ldots,t_B$,
  leading to data $\nvec = (n_1,n_2,\ldots,n_B,n_{B+1})$, where $n_i$ is the
  number of cells that die in time bin $(t_{i-1},t_i]$, with $\sum_in_i=m$,
  $t_0=0$ and $t_{B+1}=\infty$;
  \item only noisy measurements of the proportion of dead cells (out
    of a large number of cells) are available at time points
    $t_1,t_2,\ldots,t_B$, giving
    $\pvec^{obs}=(p^{obs}_1,p^{obs}_2,\ldots,p^{obs}_B)$, where
    $p^{obs}_i$ is the observed proportion of dead cells at
    time~$t_i$.
\end{enumerate}
To simplify what follows, we will assume that the initial population
level in each cell is known to be~$x_0=10$. Note that the methods we
describe in this section make use of the analytic expressions
\eqref{eq:bd_cdf} and \eqref{eq:p0}.

\subsection{Inference using known death times}
\label{sec:bd_case1}
If the times of each cell death are known, the likelihood is
$\pi(\tvec^e|\lambda,\mu)=\prod_{i=1}^m p_0(t_i^e|\lambda,\mu)$,
where $p_0(\cdot)$ is as in \eqref{eq:p0}. Therefore, by Bayes
Theorem, the posterior density is given by
\begin{equation*} 
\pi(\lambda,\mu|\tvec^e)\propto\pi(\lambda,\mu)\,\pi(\tvec^e|\lambda,\mu)
\end{equation*}
where $\pi(\lambda,\mu)$ is the prior density for $(\lambda,\mu)$.
This posterior distribution is non-standard but can be targeted using
a simple Metropolis-Hastings scheme which uses a joint update
consisting of (independent) random walks (on a log scale) for each
parameter. Such a scheme accepts proposal $(\lambda^*,\mu^*)$ with
acceptance probability $\min(1,A)$, where
\[
A=\frac{\lambda^*\mu^*\pi(\lambda^*,\mu^*)}{\lambda\,\mu\,\pi(\lambda,\mu)}\times
\prod_{i=1}^m\frac{p_0(t_i^e|\lambda^*,\mu^*)}{p_0(t_i^e|\lambda,\mu)}.
\]
Here the additional term $\lambda^*\mu^*/(\lambda\mu)$ results from
the log-normal proposal ratio
$q(\lambda,\mu|\lambda^*,\mu^*)/q(\lambda^*,\mu^*|\lambda,\mu)$.

\subsection{Inference using cell census data}
\label{sec:bd_case2}
Now suppose that the exact cell death times are not observed and
instead only the dead-alive status of each cell is observed at a
series of census times $t_1,t_2,\ldots,t_B$. From this information we
can determine the number of cells $n_i$ that die in
$(t_{i-1},t_i]$, $i=1,\ldots,B+1$. The likelihood is now
  $\pi(\nvec|\tvec,\lambda,\mu)=\prod_{i=1}^{B+1}
  \{P_0(t_i|\lambda,\mu)-P_0(t_{i-1}|\lambda,\mu)\}^{n_i}$, where
  $t_0\equiv 0$ and $P_0(\cdot)$ is as in \eqref{eq:bd_cdf}.  As
  before, realisations can be simulated from the posterior
  distribution, now given by
\begin{equation*} 
\pi(\lambda,\mu|\nvec,\tvec)\propto\pi(\lambda,\mu)\,\pi(\nvec|\tvec,\lambda,\mu),
\end{equation*}
using joint independent random walk proposals (on a log scale). Here
proposals $(\lambda^*,\mu^*)$ are accepted with probability
$\min(1,A)$, where
\[
A=\frac{\lambda^*\mu^*\pi(\lambda^*,\mu^*)}{\lambda\,\mu\,\pi(\lambda,\mu)}\times      
\prod_{i=1}^{B+1}\left\{\frac{P_0(t_i|\lambda^*,\mu^*)-P_0(t_{i-1}|\lambda^*,\mu^*)}
        {P_0(t_i|\lambda,\mu)-P_0(t_{i-1}|\lambda,\mu)}\right\}^{n_i}.
\]

\subsection{Inference using noisy measurements of cell death proportions}
\label{sec:bd_case3}
A more typical experimental scenario is one where we cannot observe
the numbers of cells that die between the census time points. Instead
all that can be observed is the proportion of cells that are dead, and
this measurement is also subject to error. We will assume an additive
normal error structure on the logit scale, that is, the observation
model is
\[
y_i=\logit\,p_{t_i}(\lambda,\mu)+\sigma\epsilon_i, 
\]
for $i=1,2,\ldots,B$, where $y_i=\logit\,p^{obs}_i$ is the logit of
the observed proportion, $p_t(\lambda,\mu)$ is the probability of a
cell being dead at time~$t$ and the $\epsilon_i$ are independent
standard normal quantites. Note that, for this simple birth-death
process, we have a closed form expression~\eqref{eq:bd_cdf} for
$p_t(\lambda,\mu)$. We will assume that $(\lambda,\mu)$ and $\sigma$
are independent \emph{a priori}, in which case the posterior density
is given by
\[
\pi(\lambda,\mu,\sigma|\yvec)
\propto\pi(\lambda,\mu)\pi(\sigma)\,\pi(\yvec|\lambda,\mu,\sigma).
\]
In this scenario, the likelihood is
\begin{equation}
\label{eq:bd_case3_like}
\pi(\yvec|\lambda,\mu,\sigma)
=\prod_{i=1}^B\phi\left\{y_i|\logit\,p_{t_i}(\lambda,\mu),\sigma^2\right\},
\end{equation}
where $\phi(\cdot|m,v)$ denotes a normal density with mean~$m$ and
variance~$v$. We can build an MCMC scheme targeting the posterior
distribution by a joint Metropolis-Hastings step with independent
symmetric normal random walk proposals (on the log scale) for
$\lambda$, $\mu$ and~$\sigma$. A proposal $(\lambda^*,\mu^*,\sigma^*)$
is accepted with probability $\min(1,A)$, where
\[
A=\frac{\lambda^*\mu^*\sigma^*\pi(\lambda^*,\mu^*)\pi(\sigma^*)}
{\lambda\,\mu\,\sigma\,\pi(\lambda,\mu)\pi(\sigma)}\times
\prod_{i=1}^B\frac{\phi\left\{y_i|\logit\,p_{t_i}(\lambda^*,\mu^*),{\sigma^*}^2\right\}}
{\phi\left\{y_i|\logit\,p_{t_i}(\lambda,\mu),\sigma^2\right\}}.
\]

\subsection{Comparison of data scenarios}
\label{sec:cmp1}
We will compare the posterior distributions under these three data
scenarios by using simulated datasets. We will simulate the cell
dynamics assuming that each cell has an initial population size
$x_0=10$ and take the birth and death rates as $\lambda=0.6$ and
$\mu=1$. In our analyses we assume that the prior distribution is not
inconsistent with the truth by taking fairly weak independent
log-normal components, with each component median set at the true
value, that is, take $\lambda\sim\LN(\log\,0.6,2)$ and
$\mu\sim\LN(0,2)$.

We base our analysis on datasets of size $n=100$ and $n=1000$
simulated under scenario~(a). In scenario~(b), the simulated data from
scenario~(a) is discretised and we take the final bin to be
$(t_B=11,t_{B+1}=\infty)$ so that the final bin contains all cells
which die after time $t=11$.  We consider the effect of different
discretisations of the datasets by fixing the census times in $(1,11]$
to be on regular grids of different size. Note that we do not consider
early time-points in $(0,1)$ as, with our choice of parameters
$(\lambda,\mu)$, the process changes very little in this time
interval.  Specifically we consider pooling the data into $B=10,25,50$
intervals, that is, look at time bins (except the final bin) with
width $t_i-t_{i-1}=10/B=1,0.4,0.2$, $i=1,\ldots,B$. We also
investigate in scenario~(c), the impact on the posterior distribution
of only observing cell death proportions at census times. Here we
simulate datasets on proportions with small, medium and large levels
of measurement error ($\sigma=0.3,0.5,0.7$). In the analysis of these
datasets we take $\sigma\sim\LN(\log\,0.5,0.5)$ as another independent
component in our prior distribution.

\begin{figure}
\centering
\includegraphics[scale=0.4]{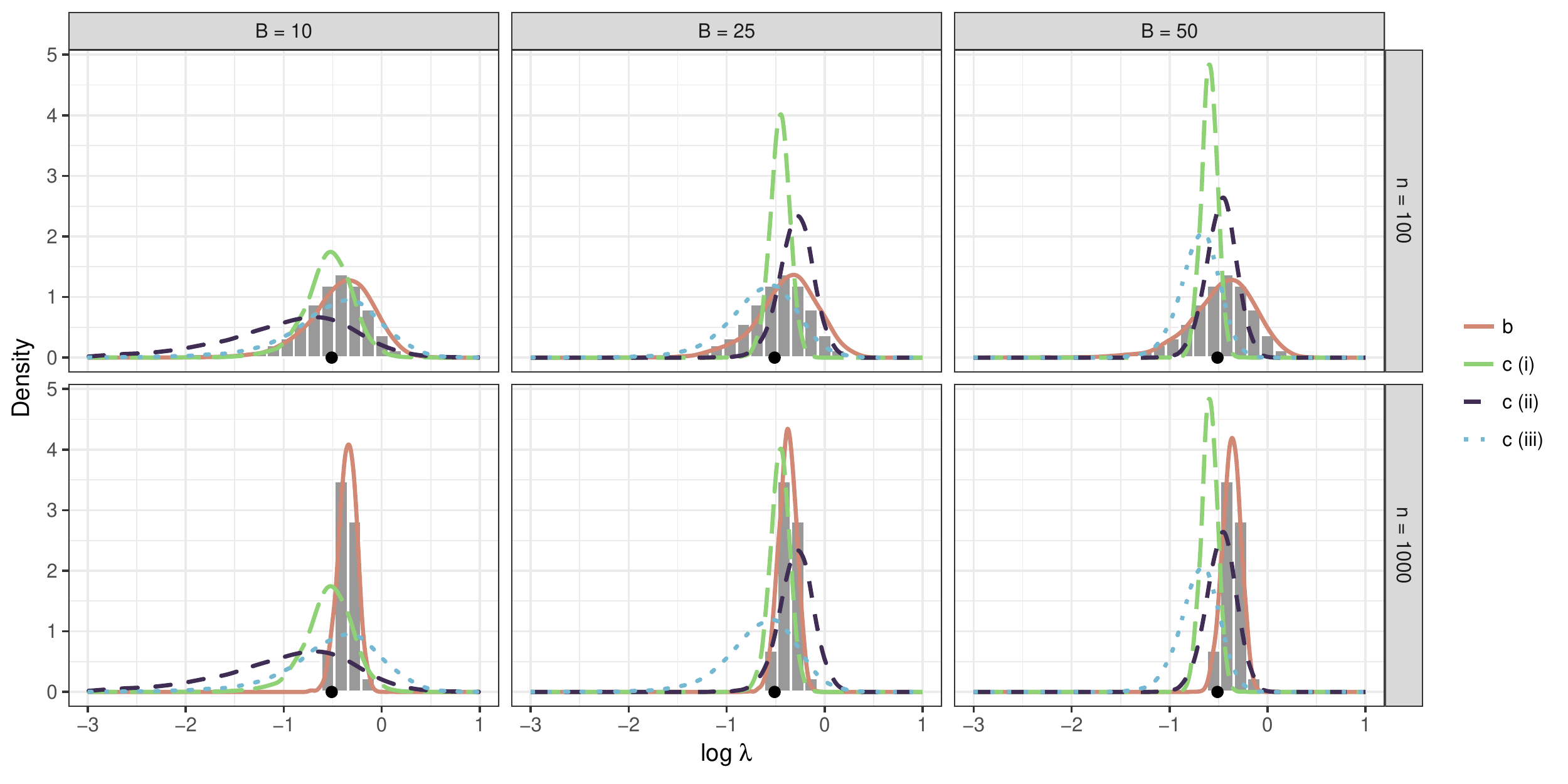}
\includegraphics[scale=0.4]{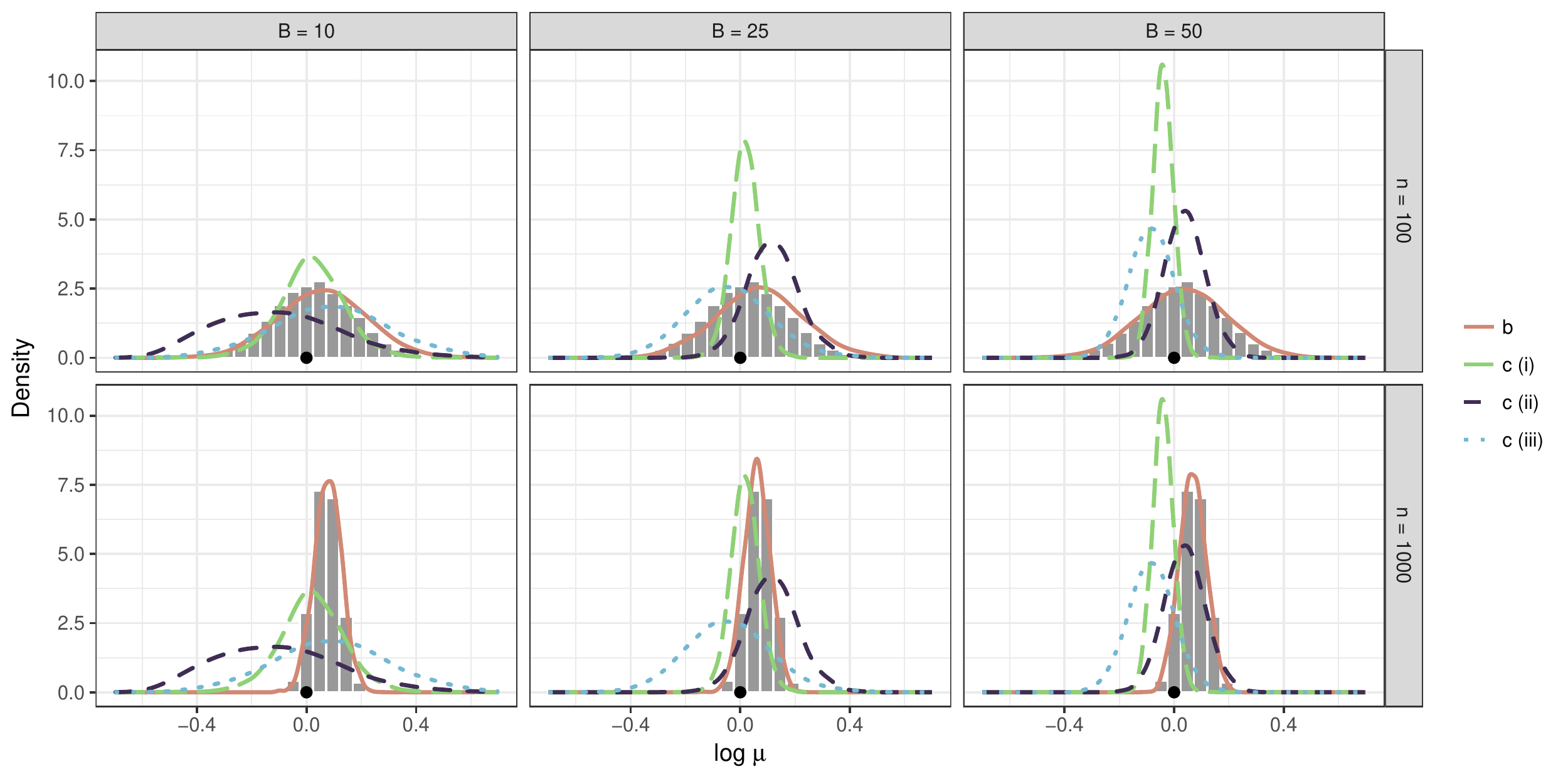}
\includegraphics[scale=0.4]{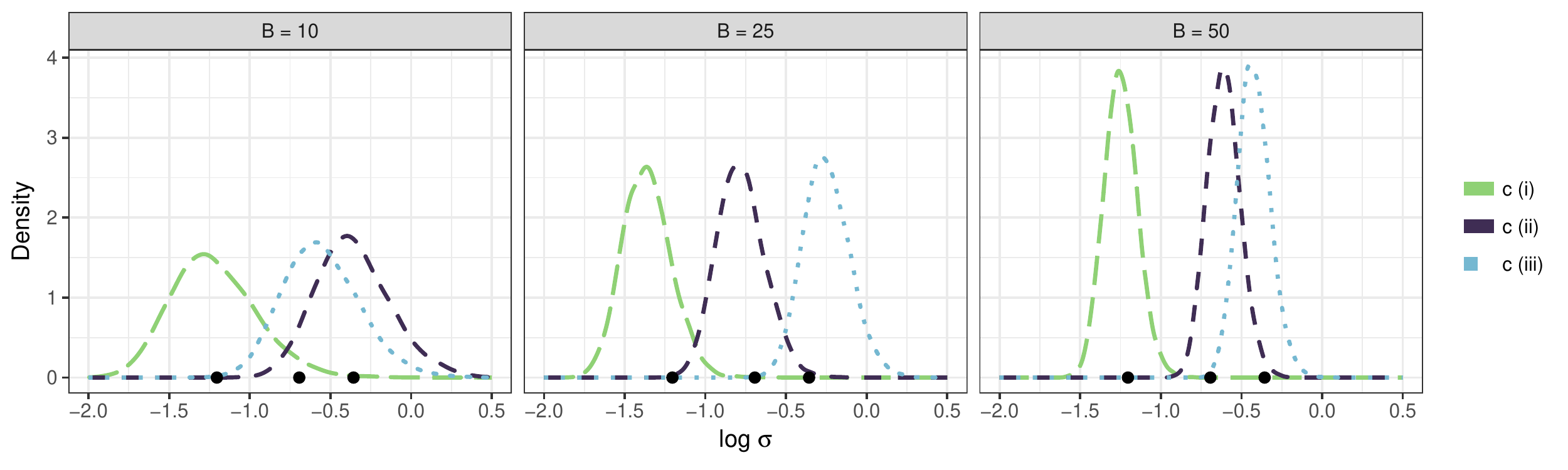}
\caption{Marginal posterior density histograms for $\log \lambda$
    (top) and $\log \mu$ (middle) using the exact algorithm and
    noise-free data from scenario (a), together with posterior
    densities for $\log \lambda$, $\log \mu$ and $\log \sigma$
    (bottom) using noise-free data from scenario (b) and noisy data
    from scenario (c), with different sizes, levels of discretisation
    ($B$) and levels of measurement error ($\sigma$).  Posterior
    densities determined by using $10^3$ realisations from MCMC
    schemes.  The true values are given by solid circles.}
\label{fig:inference1} 
\end{figure}

We now examine the effect of these various data scenarios on posterior
inference. In all cases, the posterior distribution has been
constructed after running MCMC schemes in which the first 100
iterations have been discarded as burnin and then the next $10^6$
realisations thinned by $10^3$ to obtain an (almost un-autocorrelated)
posterior sample of size $10^3$. Figure~\ref{fig:inference1} shows the
marginal posterior densities for model parameters $\log\lambda$,
$\log\mu$ and $\log\sigma$ under scenarios (a), (b) and (c) described
above.  It is clear that, under scenarios (a) and (b), the level of
discretisation in the data has very little effect on the posterior
distribution, even for the most coarse discretisation ($B=10$).  In
general, the marginal posterior distributions under scenario~(c) have
greater precision than those under scenarios~(a) and~(b).  Finally,
and unsurprisingly, under scenario~(c), posterior uncertainty for the
model parameters increases as the level of noise~($\sigma$) increases
in the data.

\section{Inference using simulator output from the underlying process}
\label{sec:simalg}
In the previous section, determining the posterior distribution using
noisy data on proportions (scenario~(c)) was made rather
straightforward because we have an analytic expression for the
probability of cell death~$p_t(\lambda,\mu)$. Unfortunately this is
generally not the case for stochastic kinetic models. Therefore we now
explore the impact of using alternative inference strategies which
rely instead on using simulated realisations from the model.

We can estimate $p_t(\lambda,\mu)$ by first simulating cell
trajectories for each of $n$ cells and then calculating the proportion
of these cells that are dead at time~$t$. Such an estimate
$\hat{p}_{t,n}(\lambda,\mu)$ has binomial sampling error since
$n\hat{p}_{t,n}(\lambda,\mu)\sim\textrm{Bin}\{n,p_t(\lambda,\mu)\}$, and is
unbiased and consistent.  As before we can build an MCMC scheme by
using a Metropolis-Hastings step with independent symmetric normal
random walk proposals (on the log scale) for $\lambda$, $\mu$
and~$\sigma$. After simulating from the model to obtain a path of
proportions
$\vec{\hat{p}}_n^*=\bigl\{\hat{p}_{t_i,n}(\lambda^*,\mu^*),
i=1,\ldots,B\bigr\}$ for some choice of $n$, we can then accept the
proposal with probability $\min(1,A)$, where
\[
A=\frac{\pi(\lambda^*)\,\pi(\mu^*)}{\pi(\lambda)\,\pi(\mu)}
\times\prod_{i=1}^B\frac{\phi\left(y_i|\elogit~\hat{p}_{t_i,n}^*,\sigma^2\right)}
{\phi\left(y_i|\elogit~\hat{p}_{t_i,n},\sigma^2\right)}
\times\frac{\lambda^*\mu^*}{\lambda\mu}
\]
and 
\[
\elogit~\hat{p}=\log \left(\frac{\hat{p}+0.5/n}{1-\hat{p}+0.5/n}\right)
\]
is the empirical logit. We use the empirical logit here as this
removes any problems with pathological cases ($\hat p=0$ or~$1$). Note
that, for finite $n$, this scheme does not target the exact posterior
distribution but it does so asymptotically as the proportion estimate
is consistent and the likelihood terms are based on the asymptotic
sampling distribution of the empirical logit.

The above scheme will work well if $n$ is very large but, in practice,
limited computing resources will result in $n$ being sufficiently
small that account needs to be taken of the sampling variation in
these estimated proportions. For large~$n$, the sampling distribution
of the empirical logit ($\elogit~\hat{p}$) is a normal distribution
with mean $\logit~p$ and variance $1/\{n\hat{p}(1-\hat{p})\}$. Thus,
taking an improper constant prior for $\logit~p$ gives its posterior
distribution as a normal distribution with mean $\elogit~\hat{p}$ and
variance $1/\{n\hat{p}(1-\hat{p})\}$. Therefore we can integrate out
posterior uncertainty about $\logit~p$ in the observation model,
modifying the likelihood to
\[
\pi(\yvec|\lambda,\mu,\sigma)=\prod_{i=1}^B 
\phi(y_{t_i}|\elogit~\hat{p}_{t_i,n},\sigma^2+1/\{n\hat{p}_{t_i,n}(1-\hat{p}_{t_i,n})\}),
\]
with consequent changes to the MCMC acceptance probability.

Incidentally, it is possible to construct a \emph{pseudo-marginal}
particle filter to target the posterior
$\pi(\lambda,\mu,\sigma|\yvec)$ exactly; see, for example,
%\cite{Ecuyer2009} and 
\cite{andrieu2010}.  We looked at schemes that
use either a Monte Carlo or a sequential Monte Carlo estimate of the
likelihood $\pi(\yvec|\lambda,\mu,\sigma)$. However we found that, for
our simple birth-death process, these schemes suffered from a much
inferior computational performance (effective sample size per cpu
second) than the scheme outlined above. This however might not be the
case in larger more complex models.

\subsection{Gaussian process emulators}
\label{sec:GPalg}
The previous analysis required that proportions of cell death be
simulated at each step of the MCMC algorithm. In all but the most
simple models, simulating from the underlying model to obtain these
proportions is far too time consuming. For example, calculating $100$
proportions, each from $n=1000$ realisations over $(0,10)$ of the
birth-death model, takes around 1 cpu sec whereas it takes around 700
cpu secs to generate the same information from the slightly larger
Schl\"{o}gl model \citep{owenWG15}, and much longer for more complex
models. In this section we consider how Gaussian process (GP)
emulators might be used to expedite inference when the simulator is
not very quick; see, for example, \cite{Rasmussen} for a background on
GP emulators. They have been used by many authors for the emulation of
complex deterministic models \citep{Kennedy2000,Kennedy2001} and for
complex stochastic models
\citep{henderson09,HendersonEtAl2010,baggaley2012}.
%\cite{henderson09}, \cite{HendersonEtAl2010} and \cite{baggaley2012}.

Determining an estimate $\hat{p}_{t,n}(\lambda,\mu)$ of the cell death
proportion at time~$t$ is too computer-intensive and so we seek to
model its sampling distribution, smoothing over $(\lambda,\mu)$-values
and accounting for binomial sampling error, using a Gaussian
process. We know that, for large~$n$, $\elogit~\hat{p}_{t,n}$ is
almost normally distributed and so we will seek a Gaussian process
emulator (approximation) for
$x_{t,n}(\lambda,\mu)=\elogit~\hat{p}_{t,n}(\lambda,\mu)$.  Note that,
because of the form of the likelihood, we do not need an emulator
across time. Rather, we need an emulator only at the time points at
which data are observed. Thus we will need to construct $B$ GP
emulators over $(\lambda,\mu)$-space. The process of fitting each GP
emulator is fairly straightforward and a major computational benefit
is that they can be fitted in parallel. 

We now describe how a GP emulator can be constructed for a particular
time point~$t$. The inputs to the GP are $\thetavec=(\lambda,\mu)$.
First we need to construct our training data, that is, determine the
value of $x_{t,n}(\thetavec)$ at a number of $\thetavec$-values. There
are many possible choices of $\thetavec$-values to use: we will use a
maximin Latin hypercube design (LHD) as these are space filling and
have been shown to be effective in other work
\citep{henderson09,baggaley2012}.  We begin by constructing an
$n_d=2000$-point LHD in $(\log\lambda,\log\mu)$ over the central
$95\%$ region of the prior distribution and then exclude any design
points that give extreme proportions, that is, proportions that are
clearly inconsistent with the data; here we exclude proportions
outside $(0.005,0.995)$. The main reason for this pragmatic step is
that when using relatively small designs, design points which have
extreme (logit) proportions can be very influential in the GP fit and
lead to GPs which fit poorly in the main area of posterior support.
The limits of the interval $(0.005,0.995)$ we report here were
determined by sequentially expanding the range from $(0.05,0.95)$
until a significant change to the GP fit was observed.  After
accounting for such deletions, this typically left around $n_d=150$
design points to use to fit the GPs.

The benefit of using a Gaussian process emulator for
$x_{t,n}(\thetavec)$ is that, as the distribution of
$x_{t,n}(\thetavec)$ at any finite collection of points
$\Thetavec=(\thetavec_i,i=1,\ldots,n_d)$ has a Gaussian distribution,
the fitted GP has a Gaussian distribution for $x_{t,n}(\thetavec^*)$
at a new point~$\thetavec^*$. This distribution has mean and variance
that depend on the prior mean function $m_t(\thetavec)$ and covariance
function $K_t(\thetavec,\thetavec')$ of the GP and the training data
$\D_t=\{(\thetavec_i,x_{t,n}(\thetavec_i)),i=1,\ldots,n_p\}$. We can
also account for the training data being estimated proportions by
adding a nugget term to the covariance function. Thus
$x_{t,n}(\thetavec^*)|\D_t\sim
N\{m_t^*(\thetavec^*),v_t^*(\thetavec^*)\}$, where
\begin{align}
\label{eq:GPmean}
m_t^*(\thetavec^*)&=m_t(\thetavec^*)+\K_t(\thetavec^*,\Thetavec)^\top
\K_t(\Thetavec,\Thetavec)^{-1}\{\yvec_t-\vec{m}_t(\Thetavec)\} 
\intertext{and}
\label{eq:GPvar}
v_t^*(\thetavec^*)&=
\K_t(\thetavec^*,\thetavec^*)
-\K_t(\thetavec^*,\Thetavec)^\top \K_t(\Thetavec,\Thetavec)^{-1}
\K_t(\Thetavec,\thetavec^*),
\intertext{where}
\K_t(\Thetavec,\Thetavec)&=
K_t(\Thetavec,\Thetavec)+\diag\bigl(n\,\eexpit\{m^*_t(\thetavec_i)\}
[1-\eexpit\{m^*_t(\thetavec_i)\}]\bigr)^{-1}\notag
\end{align}
and $\eexpit$ is the inverse of the empirical logit, that is,
$\eexpit(m)=\{e^m(1+0.5/n)-0.5/n\}/(1+e^m)$.

Inspection of the training data $\D_t$ shows that a mean function
which includes linear and quadratic terms in $\log\lambda$ and
$\log\mu$ will capture most of the dependence on the inputs. We chose
to estimate the parameters in this function using least squares. An
alternative might be to take a fully Bayesian approach and perhaps
assign weak prior information to these parameters. However, as the
number of training points~$n_d$ is reasonably large, this fully
Bayesian approach typically results in the fitted process being a
Student-$t$ process with a large number of degrees of freedom; see
\cite{pmlr-v33-shah14}. This is a more complicated process but one
which is very close to a Gaussian process.  We therefore choose to
ignore posterior uncertainty in the parameters of the mean function
and use the more straightforward Gaussian process.  Note that this
(simple) approach essentially fits a zero mean Gaussian process to the
residuals from the least squares fit. Thus we take mean function
\[
m_t(\thetavec)=b_{0t}+b_{1t}\log\lambda+b_{2t}\log\mu +
b_{3t}(\log\lambda)^2+b_{4t}(\log\mu)^2 + b_{5t}(\log\lambda)(\log\mu),
\]
where the $b_{it}$ are the least squares estimates. We use a Gaussian
covariance function
\[
K_t(\thetavec_i,\thetavec_j|a_t,\rvec_t)
=a_t\exp\left\{-\frac{(\log\lambda_i-\log\lambda_j)^2}{r_{1t}^2}  
-\frac{(\log\mu_i-\log\mu_j)^2}{r_{2t}^2}\right\}
\]
which has a variance parameter~$a_t$ and correlation length parameters
$\rvec_t=(r_{1t},r_{2t})$ and we assign fairly weak independent
log-normal $\LN(0,10)$ priors to these parameters. Their posterior
density is given by
\[
\pi(a_t,\rvec_t|\D_t)\propto\pi(a_t,\rvec_t)\,\pi(\D_t|a_t,\rvec_t),
\]
where the likelihood term $\pi(\D_t|a_t,\rvec_t)$ is an
$n_d$-dimensional normal density with mean $m_t(\Thetavec)$ and
covariance matrix $\K_t(\Thetavec,\Thetavec)$. Realisations from this
posterior can be obtained via a Metropolis-Hastings algorithm with
(independent) symmetric random walk proposals (on the log
scale). Strictly speaking the fitted GP should be
$E\{x_{t,n}(\thetavec^*)|\D_t\}$, where the expectation is taken with
respect to the posterior distribution of the GP parameters
$(a_t,\rvec_t)$. However, like many authors, we found very little
difference between this fitted GP and its delta approximation, that
is, the fitted GP evaluated at the posterior mean of its parameters
\citep{henderson09,baggaley2012}.

Recall that the benefit of using a Gaussian process emulator for
  $x_{t,n}(\lambda,\mu)$ is that the fitted GP at a new
  point~$(\lambda^*,\mu^*)$ has a Gaussian distribution, with
  $x_{t,n}(\lambda^*,\mu^*)|\D_t\sim
  N\{m_t^*(\lambda^*,\mu^*),v_t^*(\lambda^*,\mu^*)\}$, where the mean
  and variance terms are as in \eqref{eq:GPmean} and \eqref{eq:GPvar}.
  We can now use these fitted GPs to approximate the distribution of
  the observed proportions and thereby approximate the likelihood
  \eqref{eq:bd_case3_like} as
\[
\pi(\yvec|\lambda,\mu,\sigma)=\prod_{i=1}^B
\phi(y_{t_i}|m_{t_i}^*(\lambda,\mu),v_{t_i}^*(\lambda,\mu)+\sigma^2\}).
\]
Therefore we can obtain a posterior sample via a Metropolis-Hastings
algorithm which uses (independent) symmetric random walks (on a log
scale) for each parameter. Note that this algorithm is very fast
compared to the previous one as there is no need to simulate
realisations from the model - this is the benefit of using GP
emulators which have been fitted off-line.

\subsubsection{Emulators with sparse covariance functions}
\label{sec:sGPalg}
Computational efficiency gains may be achieved if the emulators are
constructed using a sparse covariance function. The idea is to take
advantage of the near sparsity of covariance matrices used in GP
calculations by constructing them in a way such that they can be
stored as sparse matrices. Here the main gain is that computationally
efficient sparse matrix algorithms can then be used to speed up
operations such as matrix inversions which would otherwise scale with
$\mathcal{O}(n_d^3)$, where $n_d$ is the number of points in GP
design~$\Thetavec$. This speed-up can be particularly beneficial when
fitting GP as such matrix inversions are required at each step of the
MCMC fitting algorithm. Also the loss in accuracy of the GP is small
when the covariance matrices used to fit the GP are nearly sparse.

\cite{Kaufman2011a} describe a sparse covariance function for a
process with input dimension~$n_p$: it has $(i,j)$th entry
\[
K(\thetavec,\thetavec')_{ij}=a\prod_{k=1}^{n_p} R_k(\Delta_{ijk};\tau_k)
\]
where $\Delta_{ijk}=|\theta_{ik}-\theta_{jk}'|$ and the correlation
function in dimension~$k$ is the Bohman function
\[
R_k(\Delta_{ijk}; \tau_k)= 
\begin{cases}
(1-\Delta_{ijk}/\tau_k)\cos(\pi\Delta_{ijk}/\tau_k)+\sin(\pi\Delta_{ijk}/ \tau_k)/\pi, & \Delta_{ijk} < \tau_k\\
0, & \text{otherwise}.
\end{cases}
\]
These functions typically look like a squared exponential function but
with the decay truncated at distance~$\tau_k$. Essentially each
hyperparameter $\tau_k$ measures the distance between two inputs in
dimension~$k$ before the output is assumed to be uncorrelated. The
algorithm begins by scaling all inputs to lie between 0 and 1, so that
the $\tau_k\in(0,1)$. The level of sparsity $s$ to be imposed on the
covariance function is specified by the user and represents a
trade-off between computational efficiency and accuracy. The level of
sparsity is induced through the prior distribution for
$\vec{\tau}=(\tau_1,\ldots,\tau_{n_p})$. This is taken to be a uniform
distribution over $\{\vec{\tau}\in
(0,1)^{n_p}:~\sum_{k=1}^{n_p}\tau_k/n_p\leq c\}$, where $c$ is chosen
to satisfy $c(2-c) = (1-s)^{1/n_p}$. Therefore, for example, taking
the sparsity level $s=0.90$ will ensure that $90\%$ of the
off-diagonal elements of $K(\thetavec,\thetavec')$ are zero. Taking
$s=0.90$ in our two dimensional problem gives a uniform prior over the
triangle $\{\vec{\tau}\in (0,1)^2:~\tau_1+\tau_2\leq 2c=0.346\}$.

Using a procedure very similar to that used to fit the non-sparse
emulator at time~$t$, we determine the posterior distribution for the
hyperparameters $(a_t,\vec\tau_t)$ by first taking the prior
distribution to have independent components, with the above prior for
$\vec\tau_t$ and $a_t\sim\textrm{LN}(0,10)$, and then fitting the GP via a
suitable MCMC scheme. As with the non-sparse emulators, the fitted
sparse GPs ignore posterior uncertainty on the hyperparameters and
simply fix them at their posterior mean.

\subsection{Effect of using non-sparse and sparse emulators on the
    posterior distribution}
\label{sec:cmp2}
We now look at the effect of using non-sparse and sparse emulators on
posterior inference for our model parameters. Here we compare the
marginal posterior distributions obtained using four different
inference schemes:
\begin{enumerate}[(a)]
\item $p_t(\lambda,\mu)$ is known (labelled `exact' in the figure)
\item $p_t(\lambda,\mu)$ is unknown and is estimated using the
  simulator (simulator)
\item $p_t(\lambda,\mu)$ is unknown and is estimated using a
  non-sparse emulator (emulator)
\item $p_t(\lambda,\mu)$ is unknown and is estimated using a
  sparse emulator (sparse), with sparsity level $s=0.90$.
\end{enumerate}
As in the previous section, data have been simulated from the
birth-death process, here using parameter values $\lambda=0.6$,
$\mu=1$ and $\sigma=0.5$, and an initial population size
$x_0=10$. Also, as before, the prior distribution has independent
components, with $\lambda\sim\LN(\log\,0.6,2)$, $\mu\sim\LN(0,2)$ and
$\sigma\sim\LN(\log\,0.5,0.5)$.

\begin{figure}
\centering
\includegraphics[scale=0.4]{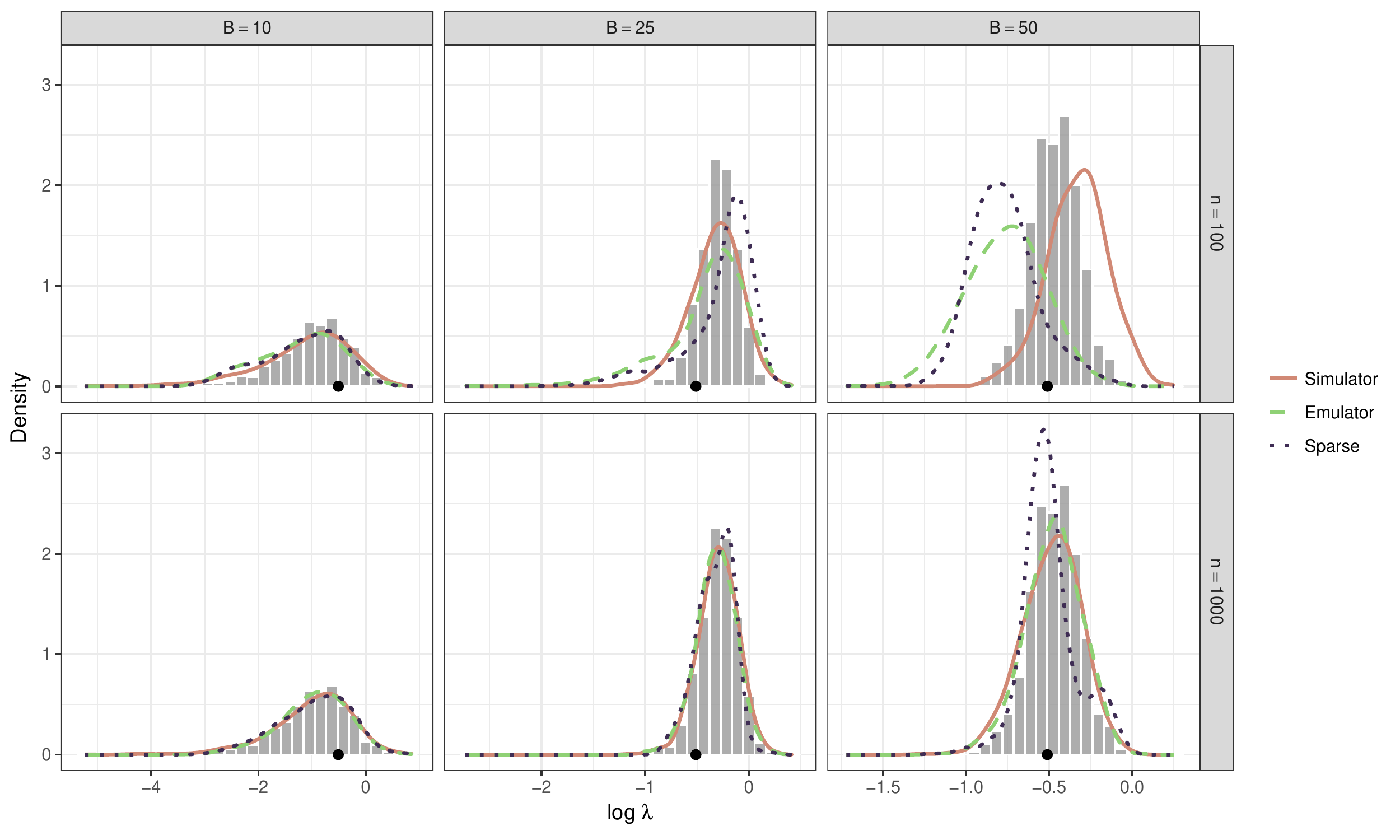}
\includegraphics[scale=0.4]{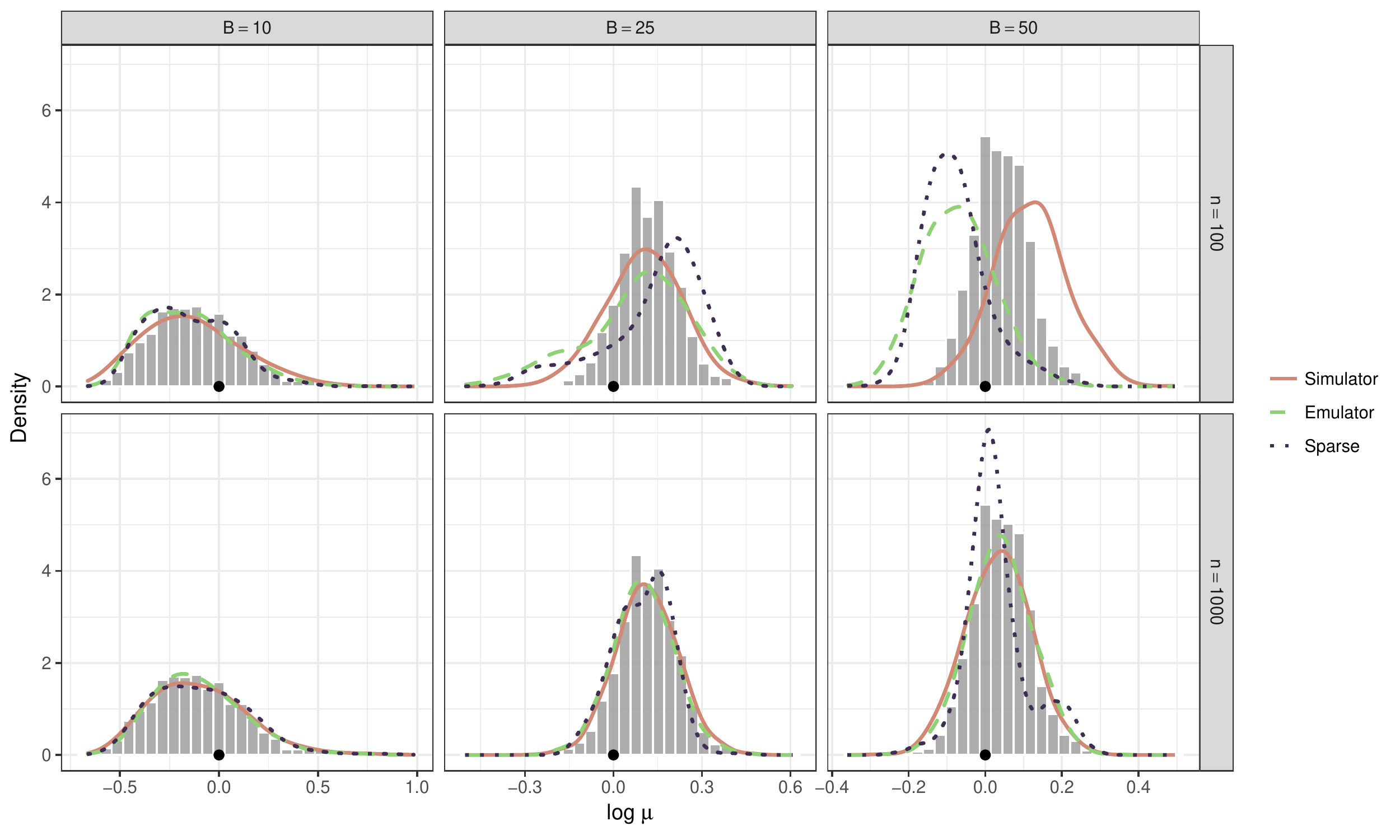}
\includegraphics[scale=0.4]{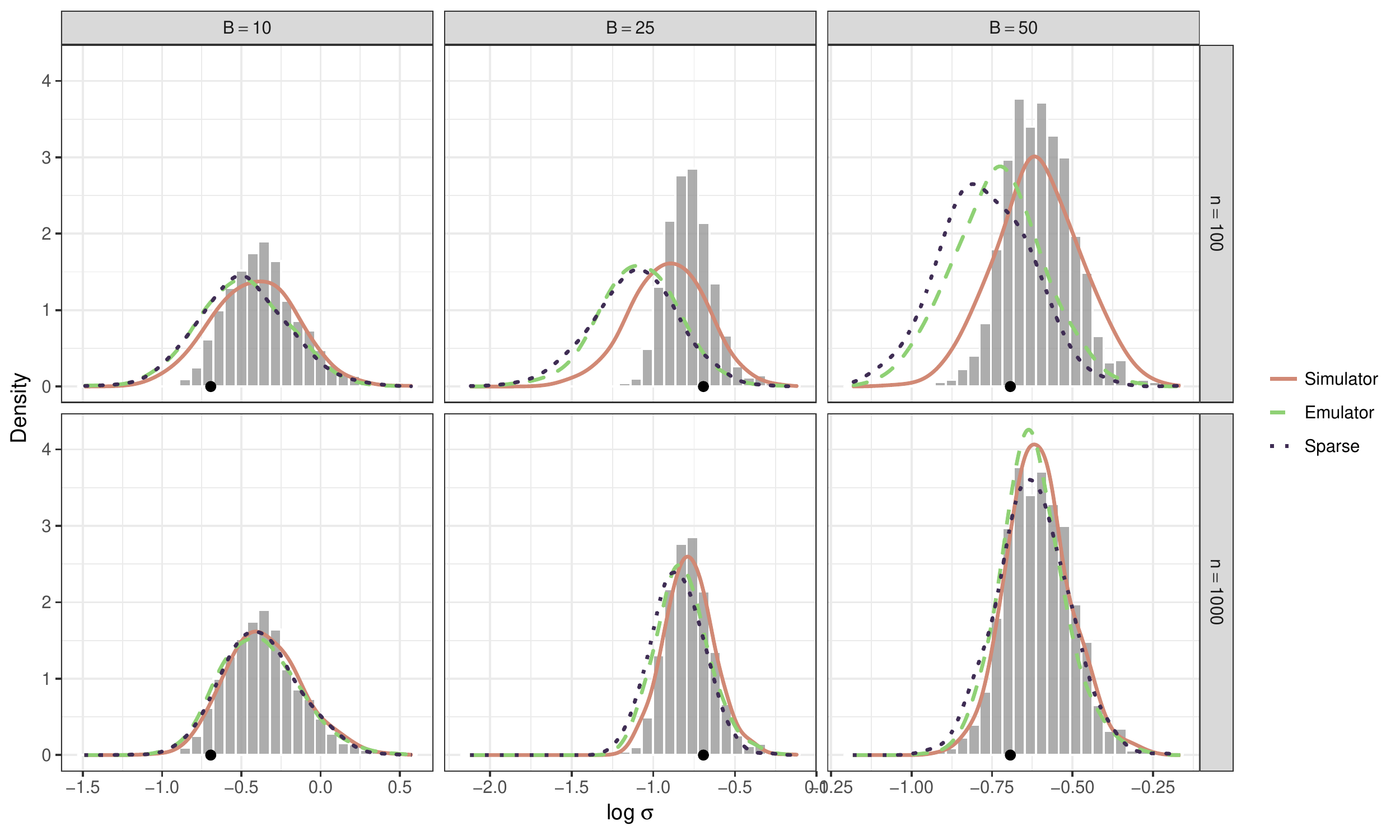}
\caption{Marginal posterior density histograms for $\lambda$
    (top), $\mu$ (middle) and $\sigma$ (bottom) using the exact
    inference scheme on noise-free data, together with posterior
    densities obtained using the simulator and non-sparse and sparse
    emulators on noisy data.  Rows refer to the number of replicates
    used in the simulator ($n$) and columns refer to the level of
    discretisation on the data ($B$). The true values are given by
    solid circles.}
\label{fig:inference2} 
\end{figure}

In all cases, the posterior distribution has been constructed after
running MCMC schemes in which the first $10^3$ iterations have been
discarded as burnin and then the next $10^4$ realisations thinned by
10 to obtain an (almost un-autocorrelated) posterior sample of size
$10^3$.  Figure~\ref{fig:inference2}
shows the marginal posterior densities for $\log\lambda$ (top),
$\log\mu$ (middle) and $\log\sigma$ (bottom). Within each panel,
columns show different levels of data discretisation ($B$) and rows
show the number of simulations ($n$) used to generate approximate
proportions $\hat{p}_{t,n}$. The figure clearly shows that regardless
of inference method, the (approximate) posterior distribution is
located in very similar regions of parameter space and that the true
parameter values are recovered well in all cases. In particular, there
is very little difference in the marginal posterior distributions when
the proportions are calculated using $n=1000$ realisations.
Unsurprisingly, the (marginal) posterior distributions obtained using
the `exact' scheme are the most precise, and those obtained by
using one of the approximate methods are fairly similar. Indeed for
$n=1000$, the emulators produce marginal posterior distributions that
are almost indistinguishable from those produced using the simulator,
with those for $n=100$ suffering from only a slight loss of precision.

\subsection{Emulator diagnostics}
Although we have seen that using emulators to determine the posterior
distribution in this simple birth-death model gives pretty accurate
results, in general it is good practice to check whether there are any
obvious discrepancies between the underlying sampling distribution of
the stochastic process at a particular time and that produced by the
emulator. There are a variety of diagnostic tools available in the
literature; see, for example, \cite{Bastos2009}. Most of these are
out-of-sample diagnostics and make use of a further set of training
data
$\D_t^\dagger=\{(\thetavec^\dagger_i,x_{t,n}^\dagger(\thetavec^\dagger_i)),i=1,\ldots,n_d^\dagger\}$
obtained by simulating from the model at a new $n_d^\dagger$-point
Latin hypercube design
$\Thetavec^\dagger=(\thetavec_i^\dagger,i=1,\ldots,n_d^\dagger)$.  One
diagnostic calculates individual prediction errors (IPE) at each point
in the LHD as
$d_t(\thetavec_i^\dagger)=\{x_{t,n}^\dagger-m_t^*(\thetavec_i^\dagger)\}/
\surd{\K_t(\thetavec_i^\dagger,\thetavec_i^\dagger)}$.  Graphical
summaries of the IPEs can be useful to assess emulator performance.
For example, if the emulator is fitting correctly then the
distribution of the IPEs should be standard normal. Large negative or
positive IPEs indicate that the emulator variance has been
underestimated. Conversely, too many very small values indicate that
the emulator variance is inflated.  An alternative way of assessing
the IPEs is to modify them using the probability integral transform
\citetext{PIT, \citealp{Gneiting2007}}. The underpinning assumptions of the GP require
that the $d(\thetavec_i^\dagger)$ should follow a standard normal
distribution. Therefore PIT statistics
$\Phi\{d(\thetavec_i^\dagger)\}$ should follow a standard uniform
distribution. It has been suggested that plots of these PIT statistics
allow departures from the GP's distributional assumptions to be
detected more easily.  Finally, an omnibus measure of the overall fit
which also accounts for the correlation between outputs can be
determined by calculating a Mahalanobis distance, here given by
\[
MD_t^2(\Thetavec^\dagger) 
= \{\xvec_{t,n}-\vec{m}_t^*(\Thetavec^\dagger)\}^\top \K(\Thetavec^\dagger,\Thetavec^\dagger)^{-1} 
\{\xvec_{t,n}-\vec{m}_t^*(\Thetavec^\dagger)\}.
\]
If the GP assumption is plausible then, given the GP parameters,
whether the emulator provides a good fit can be assessed by comparing
its value with a $\chi^2$-distribution with $n_d^\dagger$ degrees of
freedom.

\begin{figure}
\centering
\includegraphics[width=0.9\textwidth]{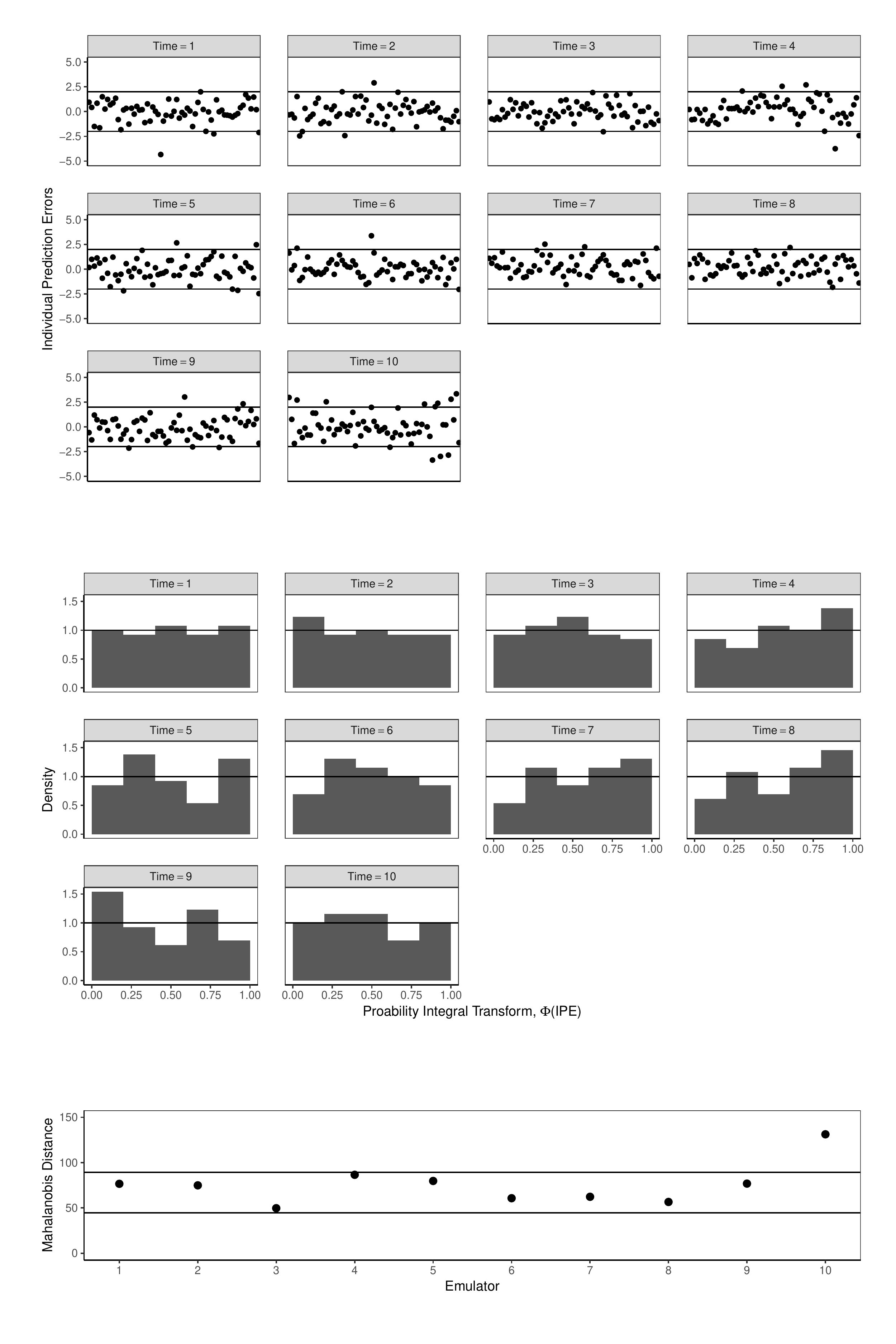}
\caption{Diagnostics for 10 non-sparse emulators (i.e. $B=10$), fitted to
  training data generated from a simulator with $n = 1000$. Individual
  prediction errors with central $95\%$ of standard normal distribution
  indicated (top), probability integral transform (middle) and Mahalanobis
  distance with central $95\%$ confidence interval of $\chi^2_{75}$ indicated
  (bottom). }
\label{fig:diagnostics_nonsparse} 
\end{figure}

\begin{figure}
\centering
\includegraphics[width=0.9\textwidth]{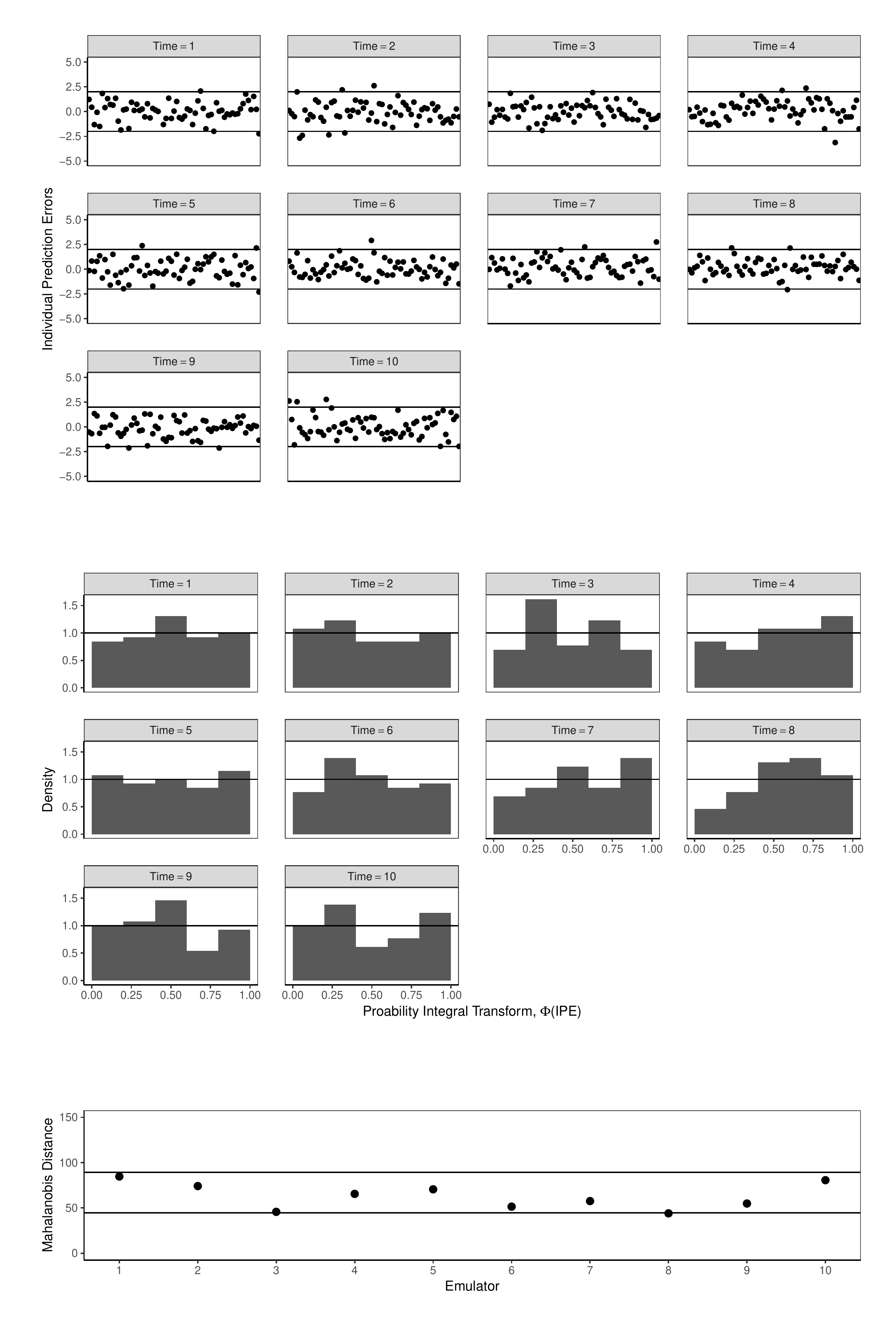}
\caption{Diagnostics for 10 sparse emulators (i.e. $B=10$), fitted to
  training data generated from a simulator with $n = 1000$. Individual
  prediction errors with central $95\%$ of standard normal
  distribution indicated (top), probability integral transform
  (middle) and Mahalanobis distance with central $95\%$ confidence
  interval of $\chi^2_{75}$ indicated (bottom). }
\label{fig:diagnostics_sparse} 
\end{figure}

Figure~\ref{fig:diagnostics_nonsparse} shows the GP diagnostics for
each of the 10 non-sparse emulators (i.e. $B=10$) fitted to training
data proportions (at $n_d^\dagger=75$ points), where each proportion
has been calculated using $n=1000$ simulator realisations. It contains
IPEs with central $95\%$ of standard normal distribution, PIT
statistics and Mahalanobis distance with central $95\%$ confidence
interval for a $\chi^2$ distribution with $n_d^\dagger=75$ degrees of
freedom. The equivalent set of diagnostic plots for the sparse
emulator are shown in Figure~\ref{fig:diagnostics_sparse}.  Both
figures are representative of the diagnostics for the other emulators
(with different~$n$ and~$B$). Overall the emulators appear to fit
fairly well and there are no unusually large IPE values. There is some
evidence of a lack of fit for some emulators, with deviation from
uniformity in the PIT histograms and the occasional slightly large
Mahalanobis distance.  The diagnostics for the sparse emulators
suggest that their fit is similar to the non-sparse emulators, though
the fit may get worse if the sparsity level were to be increased to
say 95\%.

\section{Conclusion}
We have examined the effect of various data-poor scenarios on the
accuracy of posterior inferences for the parameters of a birth-death
process. This simple process is used to describe the time evolution of
the alive-status of individual cells. Cell death is assumed to occur
when its internal population (as described by the birth-death
process) becomes extinct.  We have focused our attention on the
implications of only being able to measure the proportions of dead
cells at certain times rather than actual counts on the underlying
cell population.

We first considered the scenario where the probability of cell death
was available as an analytic expression, and considered three
data-poor scenarios. Typically analytic expressions are not available
in realistic models but, as it is available for the birth-death
process, this gives us a benchmark posterior distribution against
which to compare those obtained via simulation-based approaches.  We
constructed inference schemes for these data-poor scenarios and noted
that observing discretised death times rather than exact death times
has little effect on the posterior distribution. However, observing
exact proportions of cell death (even with a modest level of noise)
leads to more precise (marginal) posterior distributions.

We then considered the more realistic scenario in which an analytic
expression is not available for the probability of cell death. Instead
we constructed inference schemes based on estimates of such
probabilities obtained by simulating many trajectories from the
underlying stochastic model. Although simulating these trajectories
for the simple birth-death process is very fast, this is not the case
for models of reasonable size and complexity, and so alternative
strategies are needed. We developed an inference scheme based on a
Gaussian process approximation (emulator) to the simulator. We also
investigate any further computational gains that might be found by
taking advantage of the near sparsity of the emulator's covariance
function.

Comparing the various approximate (marginal) posterior distributions
with those obtained using the analytic expression for the probability
of cell death shows that the approximate methods all produce
reasonably accurate posteriors. In all cases, the (approximate)
posterior distribution is located in very similar regions of parameter
space and the true parameter values are recovered well. In particular,
there is very little difference in the marginal posterior
distributions when the proportions are calculated using $n=1000$
realisations.

We conclude with a general discussion of when, within the scenarios we
discuss in this paper, Gaussian process approximations can be usefully
employed in inference algorithms. GP-based inference algorithms can be
much more efficient if the time taken to fit them is relatively quick
compared to generating realisations using a slow simulator.  In this
paper the underlying model of cellular death is governed by the
birth-death model and as realisatons from this model are very quick to
simulate, it is never more efficient to use a GP-based algorithm.
However, with more complex models which take longer to simulate,
considerable gains can be found by employing GP approximations. For
example, in our simulation-based algorithms we need to simulate $n$
model realisations over a time interval $(0,T)$ to obtain each
proportion~$\widehat{p}_{t,n}$.  Suppose this typically takes $n\tau
T$ cpu units, where $\tau T$ is the time to simulate a $(0,T)$
realisation.  Then running the inference algorithm using the simulator
(as described in section~\ref{sec:simalg}) for $N_{iter}$ iterations
takes $n\tau TN_{iter}$ cpu units. In contrast, generating the
proportions at all of the $n_d$ training points (needed to train the
GPs) will take $n_dn\tau T$ cpu units. The main computational expense
of fitting a GP to the proportions (at a particular time-point) is in
inverting an $n_d\times n_d$ matrix at each iteration of the MCMC
fitting algorithm. This task is typically $\mathcal{O}(n_d^3)$, and so
using $N_{iter}^{GP}$ iterations to fit GPs at all $T$ time-points
will take roughly $Tn_d^3N_{iter}^{GP}$ cpu units.  Finally, running
the inference algorithm using the $T$~fitted GPs (as described in
section~\ref{sec:GPalg}) for $N_{iter}^{GPfit}$ iterations will take
around $Tn_d^3N_{iter}^{GPfit}$ cpu units. Thus in total the GP-based
inference algorithm will take $Tn_d\{n\tau
+n_d^2(N_{iter}^{GP}+N_{iter}^{GPfit})\}$ cpu units.  Further gains
can be found by employing sparse GP approximations (as described in
section~\ref{sec:sGPalg}), though \cite{Kaufman2011a} do not provide
any measure of improvement that depends on the sparsity level~$s$.  We
have found that the number of iterations needed in each inference
algorithm is very similar - this shouldn't be surprising as each
simulator/emulator is approximating the same sampling distribution at
each time-point - and so $N_{iter}\simeq N_{iter}^{GPfit}$ for both
GP-based algorithms. In conclusion, using these rough scalings, a
GP-based algorithm will be more efficient if
$n_d^3(N_{iter}^{GP}+N_{iter})<n\tau (N_{iter}-n_d)$ and, of course,
this will be true when the time $\tau T$ to simulate a $(0,T)$
realisation is reasonably large.

\bibliography{references.bib}

\end{document}